**Novel enzymes for biodegradation of polycyclic aromatic hydrocarbons: metagenomics-linked identification followed by functional analysis**

Kinga K. Nagy[1,2,a], Kristóf Takács[3,a], Imre Németh[1,a], Bálint Varga[3], Vince Grolmusz[3,4], Mónika Molnár[1,*], Beáta G. Vértessy[1,2,*]

[a]joint first authors

*corresponding authors: molnar.monika@vbk.bme.hu, vertessy.beata@ttk.hu

[1]Department of Applied Biotechnology and Food Science, Faculty of Chemical Technology and Biotechnology, Budapest University of Technology and Economics, Műegyetem rkp. 3., H-1111 Budapest, Hungary

[2]Institute of Enzymology, Research Centre for Natural Sciences, Magyar tudósok körútja 2., H-1117, Budapest, Hungary

[3]PIT Bioinformatics Group, Eötvös Loránd University, H-1117 Budapest

[4]Uratim Ltd, H-1118 Budapest

## ABSTRACT

Polycyclic aromatic hydrocarbons (PAHs) are highly toxic, carcinogenic substances. On soils contaminated with PAHs, crop cultivation, animal husbandry and even the survival of microflora in the soil are greatly perturbed, depending on the degree of contamination. Most microorganisms cannot tolerate PAH-contaminated soils, however, some microbial strains can adapt to these harsh conditions and survive on contaminated soils. Analysis of the metagenomes of contaminated environmental samples may lead to discovery of PAH-degrading enzymes suitable for green biotechnology methodologies ranging from biocatalysis to pollution control.

In the present study, our goal was to apply a metagenomic data search to identify efficient novel enzymes in remediation of PAH-contaminated soils. The metagenomic hits were further analyzed using a set of bioinformatics tools to select protein sequences predicted to encode well-folded soluble enzymes. Three novel enzymes (two dioxygenases and one peroxidase) were cloned and used in soil remediation microcosms experiments. The novel enzymes were found to be efficient for degradation of naphthalene and phenanthrene. Adding the inorganic oxidant $CaO_2$ further increased the degrading potential of the novel enzymes for anthracene and pyrene. We conclude that metagenome mining paired with bioinformatic predictions, structural modelling and functional assays constitutes a powerful approach towards novel enzymes for soil remediation.

KEYWORDS

Biodegradation, bioinformatics, metagenomic data search, novel enzymes, polycyclic aromatic hydrocarbons, soil

# 1 INTRODUCTION

Polycyclic aromatic hydrocarbons (PAHs) consisting of two or more aromatic rings are hydrophobic, semi-volatile organic compounds that may enter the environment from both natural sources and anthropogenic activities, including volcanic eruptions, forest fires and agricultural burning, incomplete combustion of organic matter, automobile exhausts, electricity-generating power plants, wood preservation, rubber tyre and cement manufacturing (Ghosal *et al.*, 2016; Samanta *et al.*, 2002). PAHs are pollutants of critical environmental and human health concern due to their low bioavailability, recalcitrance, ecotoxicity, genotoxic or carcinogenic properties (Balmer *et al.*, 2019; Kuppusamy *et al.*, 2017; Kuppusamy *et al.*, 2020). Furthermore, PAHs have been enlisted as priority environmental pollutants based on their toxicity, potential for human exposure and frequency at polluted sites (ATSDR, 2019; Keith, 2015; US EPA, 2014). Considering the toxicity and global prevalence of PAHs, the primary focus of environmental risk management activities has been to develop and apply efficient methods to remediate PAH polluted sites. Amongst the risk reduction methodologies, biodegradation based treatment (bioremediation) is emerging as an efficient and environmental-friendly option which employs microorganisms or enzymes for risk mitigation (Alegbeleye *et al.*, 2017; Gupte *et al.*, 2016; Sakshi *et al.*, 2019), whereas physical and chemical treatment methods are usually costly and highly chemical- or energy-intensive.

Numerous microorganisms (bacteria, fungi, algae) have the potential to transform or degrade PAH contaminants, among which, bacteria- and fungi-facilitated biodegradation has been studied most widely (Ghosal *et al.*, 2016). However, enzyme-based bioremediation technologies offer several advantages (i.e. greater specificity and higher mobility of the enzymes, non-dependency on expensive coenzymes or cofactors of enzymatic bioremediation) over the use of bacteria or fungi (Acevedo *et al.*, 2010; Eibes *et al.*, 2015; Villaverde *et al.*, 2019). In addition, these technologies can operate efficiently over a broader range of environmental parameters such as pH, temperature and ionic strength (Eibes *et al.*, 2015).

In the most common microbial degradation pathways, the initiating important steps are usually catalysed by redox enzymes, such as mono- and dioxygenases and peroxidases, as shown in Figure 1.

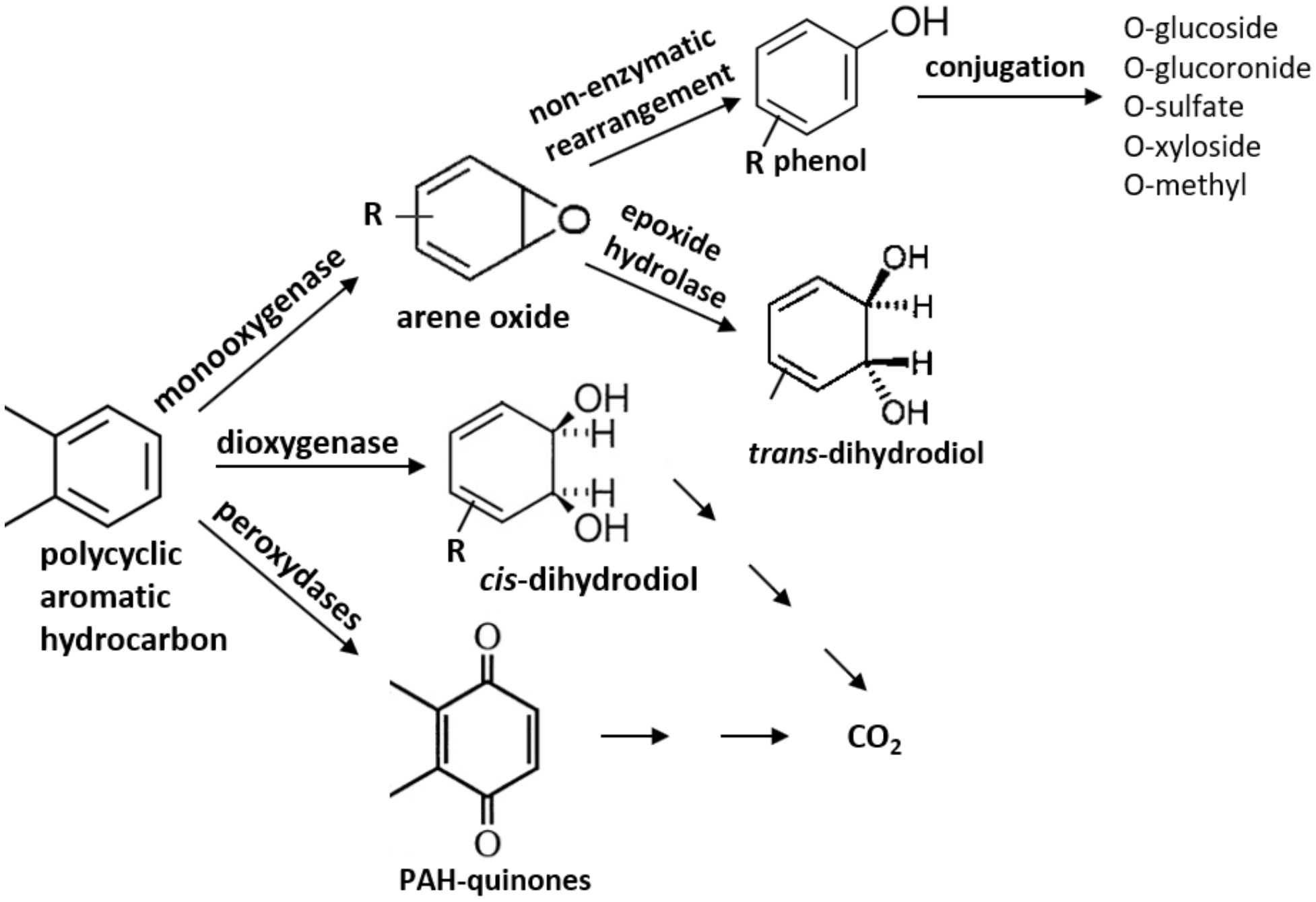

**FIGURE 1 Microbial pathways for PAH degradation**

The major enzymatic pathways leading to degradation of different PAH compounds are shown (based on Alegbeleye *et al.*, 2017; Gupte *et al.*, 2016; Sakshi *et al.*, 2019)

Ring-hydroxylating dioxygenases catalysing the initial oxidation are widely known enzymes having significant role in the biodegradation of polycyclic aromatic hydrocarbons (Jouanneau *et al.*, 2006; Jouanneau *et al.*, 2014). Dioxygenases are nonheme iron-containing enzymes, catalyze dioxygen incorporation into various organic compounds and play a key role in the complex degradation pathway of mono- and polycyclic aromatic and hetero-aromatic compounds. Dioxygenases catalyze the incorporation of both atoms of $O_2$ into substrates. All gentisate dioxygenases (GDOs) characterized to date belong to the bicupin family (Fetzner, 2012). Different peroxidase enzymes with PAH pollutants removal ability were also detected (Eibes *et al.*, 2015). The application of enzymes for degradation of pollutants demands some requirements regarding the enzyme properties: increased stability, substrate specificity, and kinetic properties (Eibes *et al.*, 2015; Rayu *et al.*, 2012). Meeting of these requirements is supported by novel engineering tools, such as metagenomics and metabolic engineering. The application of genomic information, gained from metagenomes of diverse environmental samples is a widely researched field today. Species, adapted to particular environmental conditions may contain enzymes, which are well-applicable in biotechnology, including chemical manufacturing and environmental pollution control.

Metagenomic samples, originated from highly polluted industrial areas may contain the genomes of microorganisms, which are successful in the decomposition of the industrial pollutants. Applying these naturally occurring enzymes in the rehabilitation of polluted industrial areas may enhance the cost and time conditions of such activities.

Coal gasification plants produced lighting gas and coking coal in the second half of the XIX$^{th}$ and the first half of the XX$^{th}$ centuries. Before the widespread application of natural gas in the last third of the XX$^{th}$ century, this lighting gas was used in urban areas worldwide. The industrial areas of those former factories are highly contaminated by the by-products of the manufacturing process, including polycyclic aromatic hydrocarbons (PAHs). E. g. one of the oldest coal gasification plants in North America was the Victoria Gas Company on the Vancouver Island, in the Rock Bay, British Columbia, Canada. The plant ceased production in 1952, but its former area is still contaminated by numerous pollutants. A large metagenomic study was performed in this area in 2005 by Carleton

University, analyzing the soil samples from the gasification plant (Meier *et al*., 2016). The short reads, identified from the DNA strands by Illumina HiSeq 2000 sequencing were deposited in the NCBI SRA archive by the accession numbers of SRX1419392 and SRX1419010.

In the present study, our major goal was identifying new PAH-decomposing enzymes from the metagenome, particularly focusing on the families of dioxygenases and peroxidases. We have used artificial intelligence tools for distilling the hidden common properties of PAH-metabolizing enzymes, and based on these properties, we have identified novel enzymes using the methodology described in Szalkai & Grolmusz, (2019).

The characterization and evaluation of PAH biodegradation in soil are much more complicated than in liquid cultures because of the heterogeneity and complexity of soil. Thus, many biotic and abiotic factors should be considered when studying soil bioremediation. Numerous studies demonstrated that the biodegradation rate of organic pollutants in soils depends on the environmental parameters, the soil characteristics, the type of microorganism and enzymes, as well as the nature and chemical structure of the PAH compound to be removed. Manipulation and optimization of these factors and the addition of biosurfactants (Bezza & Chirwa, 2017; Zang *et al.*, 2021), nanoparticles (Abbasi *et al.*, 2014; Parthipan *et al.*, 2022) and peroxides (Wang *et al.*, 2021; Tang *et al*., 2022) may increase biodegradation efficiency. Therefore, to design and develop a bioremediation technology, a number of factors are to be taken into account.

For several representatives of the newly identified enzymes, we have also performed functional analysis. The objectives of the functionality study were: (1) to assess the efficiency of the novel enzymes for degradation of PAH pollutants in soils, (2) to investigate the effect of calcium peroxide ($CaO_2$) as an additive, whether it can enhance the degradation, (3) to evaluate the microbial activity after bioremediation in soil. Our data indicate that these novel enzymes can efficiently degrade most of the PAH pollutants in soil microcosms, especially the low molecular weight PAHs (LMW PAHs) (e.g. naphthalene, anthracene, phenanthrene), and even some high molecular weight PAH (HMW PAH) contaminants (e.g. pyrene) and a cost-effective inorganic oxidant may increase the efficiency of the degradation. Our study presents a pipeline as a useful resource for identifying potent novel enzymes for PAH biodegradation in soils.

# 2 MATERIALS AND METHODS

## 2.1 Metagenomic analysis

In the metagenomic sequence search, we needed to solve the following problem: based on the already identified PAH-decomposing enzymes, new, still unknown, possibly better enzymes needed to be found. If we had filtered according to sequence homology with known PAH-decomposing enzymes, then we would have identified from the metagenome only slightly different "new" enzymes, relative to the already known ones. Consequently, we have chosen a more subtle, artificial intelligence-based method, as follows:

(i) First, we identified six potentially PAH-degrading enzyme groups, from database- and literature search. The members of these classes are listed in Supplementary Tables S1-S6.

(ii) Next, multiple alignments were performed for each of the six classes separately, using the Clustal Omega software (Sievers *et al.*, 2011; McWilliam *et al.*, 2013; Li *et al.*, 2015) with default parameters.

(iii) Next, we used the `hmmbuild` tool (Johnson *et al.*, 2010; Eddy, 2011) for constructing six hidden Markov models (HMM), one for each class, using the Clustal-aligned structures from step (ii). These models made possible the search for unknown sequences, having the common properties of the proteins in the starting classes.

(iv) The `hmmsearch` program (Johnson *et al.*, 2010; Eddy, 2011) found the short reads of the metagenome with the highest similarity of the HMM profiles, created in (iii). The selected short reads are listed with E-values increasing; we applied a cut-off value 0.01, that is, only those short reads were returned, which have had an E-value of 0.01 or less.

(v) The identified short reads were assembled into the longest possible sequences by using the Megahit program (Li *et al.*, 2015; Li *et al.*, 2016). Only those results were retained which had both start- and stop codons.

(vi) The identified protein sequences were compared by the NCBI BLAST program against the RefSeq non-redundant protein sequences. The hits, which have very high similarity with the known proteins were discarded as not novel.

The description of the `hmmbuild` and `hmmsearch` components of the HMMER3 suite is at the site http://hmmer.org/documentation.html.

The amino acid sequences of the hits are listed in Supplementary Table S7.

## 2.2 Bioinformatic analysis of hit sequences

In order to identify potentially well folded enzymes, the hit sequences were analyzed by several bioinformatic predictors, such as PROSO II (Smialowski *et al.*, 2012), Protein-sol (Hebditch *et al.*, 2017), ESPRESSO (Hirose & Noguchi, 2013), MEMEX (Martin-Galiano *et al.*, 2008), CRYSTALP2 (Kurgan *et al*., 2009), IUPRED (Erdős & Dosztányi, 2020). Hit protein sequences were classified according to solubility, expressability, crystallizability and flexibility. The sequences were further analyzed by multiple sequence alignment with ClustalW.

The presence of the amino acids which are essential in catalytic activity was important during selection of hits for further detailed analysis. Three-dimensional models of several hit proteins were built with SWISS-MODEL server (Waterhouse *et al.*, 2018).

### 2.3 Cloning, expression and purification

For protein expression the relevant protein coding genes were obtained by gene synthesis, sequences were ordered from GenScript, cloned into plasmid pET15b. This plasmid encodes an amino- terminally His-tagged variant of the enzymes. *E. coli* Rosetta cells with the recombinant pET15b were grown at 37 °C overnight in 5ml LB-media with chloramphenicol (34 µg/ml) and carbenicillin/ampicillin (50 µg/ml) then transferred into 500 ml fresh LB media. The cells were grown to an OD600 nm of 0.8 and the expression of the proteins induced by adding isopropyl-1-thio-b-D-galactopyrano-side (IPTG) (0.7 mM). The protein expression took place at 25 °C. The cells were harvested about 5 h after IPTG addition by centrifugation (4000 rpm, 20 min, 4 °C).

For purification of enzyme proteins, *E. coli* cells were resuspended into lysis buffer (Tris/HCl (50 mM, pH 8.0), 300 mM NaCl) and disrupted by using sonication. Intact cells and cell fragments were removed by centrifugation (11000 rpm, 30 min, 4°C). The enzymes were purified from the crude extracts by using a nickel affinity chromatography. The proteins were eluted by using buffers containing 50 mM HEPES (pH 7.5), 300 mM KCl, 5 mM β-mercaptoethanol and 250 mM imidazole. The solutions were dialyzed into 140 mM NaCl, 30 mM Tris/HCl (pH 7.5) containing buffer.

### 2.4 Soil microcosms to study the functionality and the efficiency of the enzymes

Laboratory soil microcosms were set up to study the functionality and efficiency of the nominated enzymes for the biodegradation of PAH compounds in soil. The 7-day long experiments were run in 27 glass reactors provided with respirometric heads and containing 200-200 g of contaminated matrix, each.

The contaminated matrix was an aleuritic sand (obtained from Elgoscar Environmental technology Plc., Hungary), artificially spiked with naphthalene, phenanthrene, anthracene and pyrene dissolved in gasoil. The spiked gasoil contamination of the soil was 8000 mg/kg. The concentrations. After spiking, the concentrations of the PAH compounds in the soil were determined by gas chromatography coupled with mass spectrometry (GC MS) using Agilent 6890 5973 N Autosampler System according to the 21470-84:2002 standard. These concentrations were 12.5 mg/kg, 25.9 mg/kg, 46.2 and 52.0 mg/kg for naphthalene, phenanthrene, anthracene and pyrene, respectively.

1 mg purified enzyme in 5 ml buffer solution as enzyme-based inoculant was added into 200 g contaminated soil. Nitrogen and phosphorous, as nutrient supply was not applied in the soil microcosms. 20 mmol $CaO_2$ was added to 12 microcosms in order to reduce bacterial activity and enhance degradation of the PAH compounds. $CaO_2$ (CAS: 78403-22-2) was purchased from Sigma-Aldrich Merck (Product No: 466271).

The experimental set-up contained one untreated control series without inoculant and calcium peroxide treatment. 20 mmol $CaO_2$ treatment was also tested without enzymes. Each treatment was run in triplicate. Table 1 shows the applied experimental treatments.

**TABLE 1. Treatment conditions of the microcosm experiments**

| Denomination | Experimental set-up |
|---|---|
| Control | 200 g contaminated soil (cont. soil) + 5 ml buffer |
| $CaO_2$ | 200 g cont. soil + 20 mmol $CaO_2$ + 5 ml buffer |
| E_99 | 200 g cont. soil + 1 mg PAH1_99 (in 5 ml buffer) |
| E_105 | 200 g cont. soil + 1 mg PAH1_105 (in 5 ml buffer) |
| E_39 | 200 g cont. soil + 1 mg PAH6_39 (in 5 ml buffer) |
| E_99+ $CaO_2$ | 200 g cont. soil + 1 mg PAH1_99 (in 5 ml buffer) + 20 mmol $CaO_2$ |
| E_105+ $CaO_2$ | 200 g cont. soil + 1 mg PAH1_105 (in 5 ml buffer) + 20 mmol $CaO_2$ |
| E_39+ $CaO_2$ | 200 g cont. soil + 1 mg PAH6_39 (in 5 ml buffer) + 20 mmol $CaO_2$ |

### *2.4.1 Methods applied to study the efficiency of the enzymes*

To determine the concentration of the contaminants in the soil as well the biological activity, samples have been collected from each microcosm after the completion of the experiments.

The concentration of PAH compounds during the experiments was determined by gas chromatography mass spectrometry (GC MS) using Agilent 6890 5973 N Autosampler System according to the 21470-84:2002 standard.

For microbial activity measurement, the Biolog EcoPlate$^{TM}$ (Biolog Inc., Hayward, CA, USA) was applied. The Biolog plate is a microtiter plate including 96 wells, which contain 31 various carbon source substrates in triplicate. These substrates are the most useful carbon sources for the physiological profiling of heterotrophic soil bacteria communities. The consumption of a certain substrate by a microbial population group results in a characteristic response pattern, thus various metabolic patterns are developed. The measurement was carried out according to Feigl *et al*., (2017). Average Well Colour Development (AWCD) was calculated for all carbon sources with the following equation:

$$AWCD = \sum \frac{(C - R)}{n}$$

Where, C is the absorbance value of the substrate-containing well, R is the absorbance of the control well, n is the substrate number. The calculated AWCD represents a quantitative measure of the general potential metabolic activity indicator of the microbial community.

### *2.4.2 Statistical analysis*

The statistical evaluation of the datasets was carried out with TIBCO Statistica™ 13.5. Software. One-way analysis of variance (ANOVA) was performed to investigate whether the treatments (enzymes, calcium peroxide or their combined application) had any effect on the examined parameter such as concentrations of PAHs and microbial activity; all p values less than 0.05 were considered statistically significant. To compare the treatments Fisher's least significant difference (LSD) test was carried out. Letters in alphabetical order are used to mark the significant effects on figures, where "a" represents the smallest average value of the data sets in all separate comparative experiments. Columns assigned with the same letter indicate that there was no significant difference between them.

# 3 RESULTS AND DISCUSSION

## 3.1 Metagenomic hits from the dioxygenase and catalase/peroxidase family of enzymes

Table 2 represents a list of hits from both the dioxygenases and catalase-peroxidase families. For the selection of hits that will be used in further detailed experimental work, we set a number of criteria. First, only full-length sequences were selected, in which the representative domains and motifs were all present as compared to homologous sequences from the NCBI database. Second, in order to identify potentially novel enzymes, we selected hit sequences that show considerable amino acid variations as compared to enzyme sequences already present in the NCBI database. The hits fulfilling these two criteria are all included in Table 2 (for hit sequences see Supplementary Table S7). The third criteria related to the crystallizability potential. Crystallizability potential was included in the criteria since it usually indicates a well-folded protein and may lead to high resolution structural insights in further studies. In this respect, we used the CrystalP2 software and removed those hits that were predicted to be non-crystallizable from the further investigations.

**TABLE 2 General properties and bioinformatic predictions of metagenomic hit enzymes**

| Identification code | Number of amino acids | BLAST first hit | % Match with BLAST's | End sign | Solubility (PROSOII) (0-1) | Solubility Protein-sol | Expressibility ESPRESSO | Expressibility (MEMEX) | Crystallizability (CRYSTALP2) |
|---|---|---|---|---|---|---|---|---|---|
| PAH1_16 | 346 | gentisate 1,2-dioxygenase [Bradyrhizobium sp. S23321] | 98.00% | 1 | yes (0.636) | 0.375 | yes | yes | non-crystallizable |
| PAH1_17 | 346 | gentisate 1,2-dioxygenase [Bradyrhizobium lablabi] | 99.00% | 1 | no (0.436) | 0.409 | yes | yes | non-crystallizable |
| **PAH1_99** | 374 | **gentisate 1,2-dioxygenase** (cupin domain-containing protein) [Acidovorax sp. OV235] | 86.00% | 1 | no (0.440) | 0.295 | yes | yes | crystallizable |
| PAH1_102 | 397 | gentisate 1,2-dioxygenase (cupin domain-containing protein) [Pseudomonas sp. GM55] | 96.00% | 1 | yes (0.685) | 0.239 | yes | yes | crystallizable |
| **PAH1_105** | 356 | **gentisate 1,2-dioxygenase** [Variovorax sp. YR216] | 91.00% | 1 | no (0.545) | 0.248 | yes | yes | crystallizable |
| PAH1_117 | 324 | gentisate 1,2-dioxygenase (cupin domain-containing protein) [Mesorhizobium australicum] | 89.00% | 1 | yes (0.620) | 0.339 | yes | yes | non-crystallizable |
| **PAH6_39** | 726 | **catalase/peroxidase** HPI [Methylotenera sp. 24-45-7] | 87.00% | 1 | no (0.518) | 0.42 | yes | yes | crystallizable |
| PAH6_78 | 719 | catalase/peroxidase HPI MULTISPECIES: [Acidovorax] | 99.00% | 1 | no (0.534) | 0.351 | yes | yes | crystallizable |
| PAH6_112 | 730 | peroxidase, partial [Lutibacter sp. BRH_c52] | 94.00% | 1 | no (0.564) | 0.467 | yes | yes | crystallizable |

Both the ESPRESSO and MEMEX servers, which estimates protein expression predicted that all of our selected protein can be expressed in *E. coli*. In further details, we focused on sequence and structural alignments to select potential novel enzymes with preserved catalytic sites as compared to previously described representatives of the dioxygenase and catalase/peroxidase enzyme families.

## 3.2 Structural analysis of the dioxygenase metagenomic hits

Figure 2A shows the alignment of the six full-length sequence hits from the dioxygenase enzyme family. Among these six sequences, only the sequences denoted as PAH1_99, PAH1_102 and PAH1_105 were predicted to be crystallizable. A close inspection of this alignment revealed an intriguing variation at a specific sequence position (highlighted in yellow on Figure 2A). At this position, the protein possesses either glycine or alanine amino acid and according to the literature this variation has a significant effect on the substrate selectivity of the dioxygenase *(Ferraroni et al.*, 2013).

If glycine is present at this position, then the enzyme can bind several substrates in the active site, while dioxygenases with alanine in this position oxidize only gentisate.

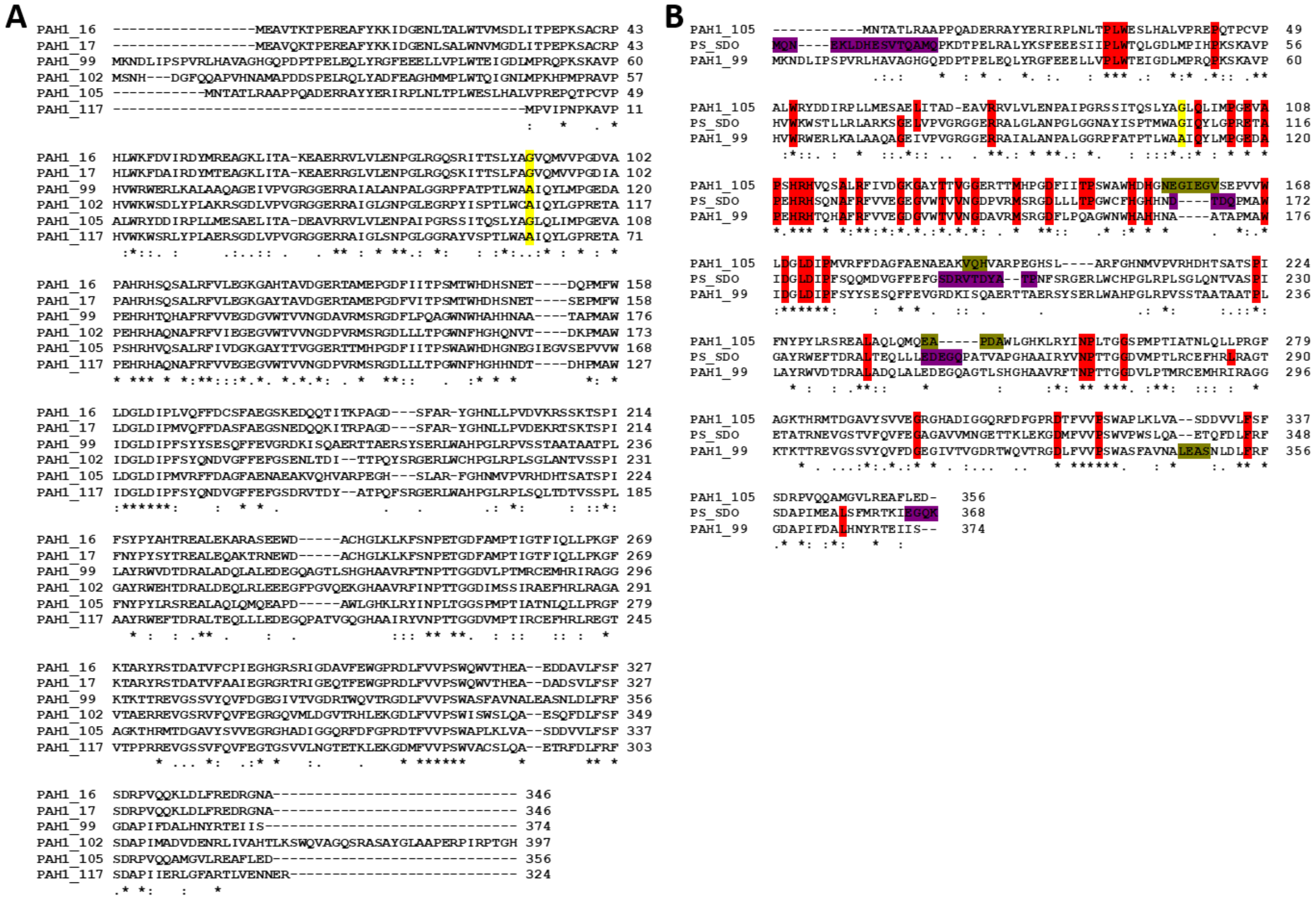

**FIGURE 2 Dioxygenase sequence alignments**.

**Panel A** Amino acid sequences of the six dioxygenases which were found during metagenomic search. The amino acids important in substrate specificity were highlighted by yellow. Glycine in this position allow several substrates binding in the active site, while dioxygenases with Alanine in this position oxidize only gentisate (Ferraroni *et al.*, 2013).

**Panel B** Amino acid sequence aligment of the two selected dioxygenases (PAH1_99 and PAH1_105) and *Pseudaminobacter salicylatoxidans* salicylate 1,2-dioxygenase. Highly conserved amino acids based on literature (Ferraroni *et al.*, 2013) labeled by red. Yellow color indicates the G/A point mutation which influences the enzyme substrate specificity. The flexible (loop) regions of the template were highlighted by purple, and the main differences in the three-dimensional structure between the template and the model were highlighted by ocher.

Our aim was to investigate one representative from both the alanine and the glycine containing variations. PAH1_105 is the single hit with a glycine residue in this critical position that was also predicted to be crystallizable, hence this sequence was selected for further characterizations. From among the two crystallizable hits with an alanine in the critical position, PAH1_99 and PAH1_102, we have selected PAH1_99 which shows more variability as compared to the sequences present in the database.

Next, we created a three-dimensional structural model of the two selected dioxygenases using the SWISS-MODEL server. In both cases the template identified in the SWISS-MODEL server was the same, namely *Pseudaminobacter salicylatoxidans* salicylate 1,2-dioxygenase (PDB ID: 3NW4). Global Model Quality Estimation (GQME) scores were 0.83 and 0.72 in case of PAH1_99 and PAH1_105, while the QMEAN scores were –1.27 and –1.47. Figure 2B shows the alignment for the two metagenomic hits and their three-dimensional template sequences: most residues strictly conserved in dioxygenase sequences based on the literature (Ferraroni *et al.*, 2013) are also conserved in the two novel metagenomic hits (cf red background).

The three-dimensional fold predicted with the SWISS-MODEL server is shown in Figure 3A (homotetramers and one subunits, respectively, PAH1_105: orange, PAH1_99: greencyan, template: yellow), while Figure 3B presents a close up from the active site, indicating several differences in amino acid positions of the two metagenomic hits. These enzymes have a homo-tetrameric structure containing a catalytic Fe(II) ion coordinated by three histidine residues in the N-terminal region (Ferraroni *et al.*, 2012). The predicted folds of the two metagenomic hits are identical with minor differences observed at one loop position (a flexible loop in the template sequence – show on Figure 2 and 3B).

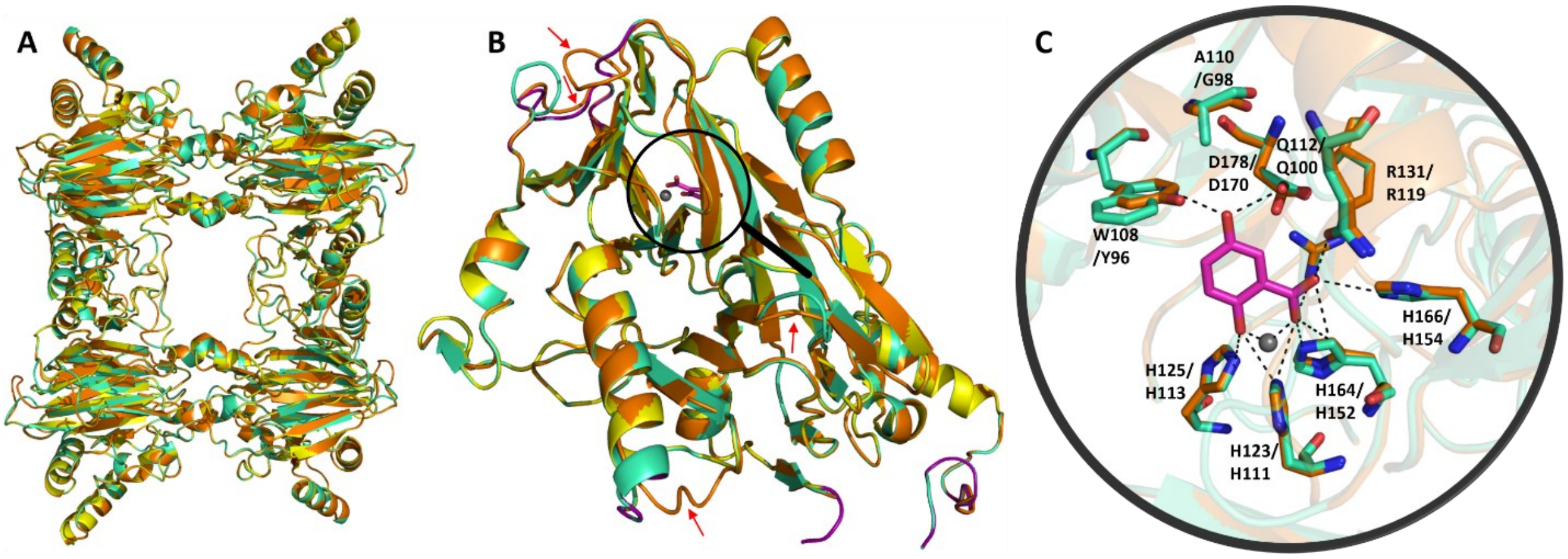

**FIGURE 3 Three-dimensional models of the two dioxygenases (PAH1_99 and PAH1_105) built based on *Pseudaminobacter salicylatoxidans* salicylate 1,2-dioxygenase (PDB ID: 3NW4).**

**Panel A** The merged homotetrameric structures of the template (yellow), and the models of PAH1_99 (greencyan) and PAH1_105 (orange)

**Panel B** A close up of a monomer with the active site. Red arrows point on the main differences in the three-dimensional structure between the template and the models (the sequences are highlighted by ocher on Fig.2B). The flexible (loop) regions of the template were colored by purple.

**Panel C** Amino acids which interact with the substrate gentisate based on *Pseudaminobacter salicylatoxidans* salicylate 1,2-dioxygenase three-dimensional structure (PDB ID: 3NW4). Colours: PAH1_99 (greencyan) and PAH1_105 (orange) with the substrate gentisate (2,5-dihydroxybenzoic acid) (magenta). The Fe(II) ion is indicated by a gray sphere. Polar contacts were labelled by black dashed lines. (First amino acid belongs to PAH1_99, the second to PAH1_105).

The wild-type *Pseudaminobacter salicylatoxidans* salicylate 1,2-dioxygenase protein sequence contains a Gly amino acid in the position 106. It was found that protein with G106A mutation oxidized only gentisate, while 1-hydroxy-2-naphthoate and salicylate were not converted. The amino acid residue Gly106 is located inside the enzyme active site cavity but does not directly interact with the substrates. In the case of the G106A mutation, based on the crystal structure of the complex, a different binding mode was observed for salicylate compared to the wild-type enzyme. The salicylate in the G106A variant coordinated to the catalytically active Fe(II) ion in an unusual and unproductive manner, since salicylate is unable to displace the hydrogen bond formed between Trp104 and Asp174 in the G106A variant. Presumably, such inefficient substrate binding may generally limit the substrate spectrum of wild type GDOs (Ferraroni *et al.*, 2013).

## 3.3 Structural analysis of the catalase-peroxidase metagenomic hits

In case of catalase-peroxidases all of the metagenomic hits were predicted as crystallizable (see Table 2), so we have selected PAH6_39 which shows more variability as compared to the sequences present in the database. Figure 4A shows the alignment of the three full-length sequence hits from the catalase-peroxidase enzyme family. A three-dimensional structural model of the selected catalase-peroxidase was also built using the SWISS-MODEL server.

The template identified in the SWISS-MODEL server was the *Synechococcus elongatus* catalase-peroxidase KatG (PDB ID: 3WNU). Global Model Quality Estimation (GQME) score was 0.91 and the QMEAN score was –1.32. Figure 4B shows the alignment for the selected catalase-peroxidase and their three-dimensional template sequences: most residues strictly conserved in catalase-peroxidase sequences according to the literature (Kamachi *et al.*, 2015) are also conserved in the novel metagenomic hits (cf red background).

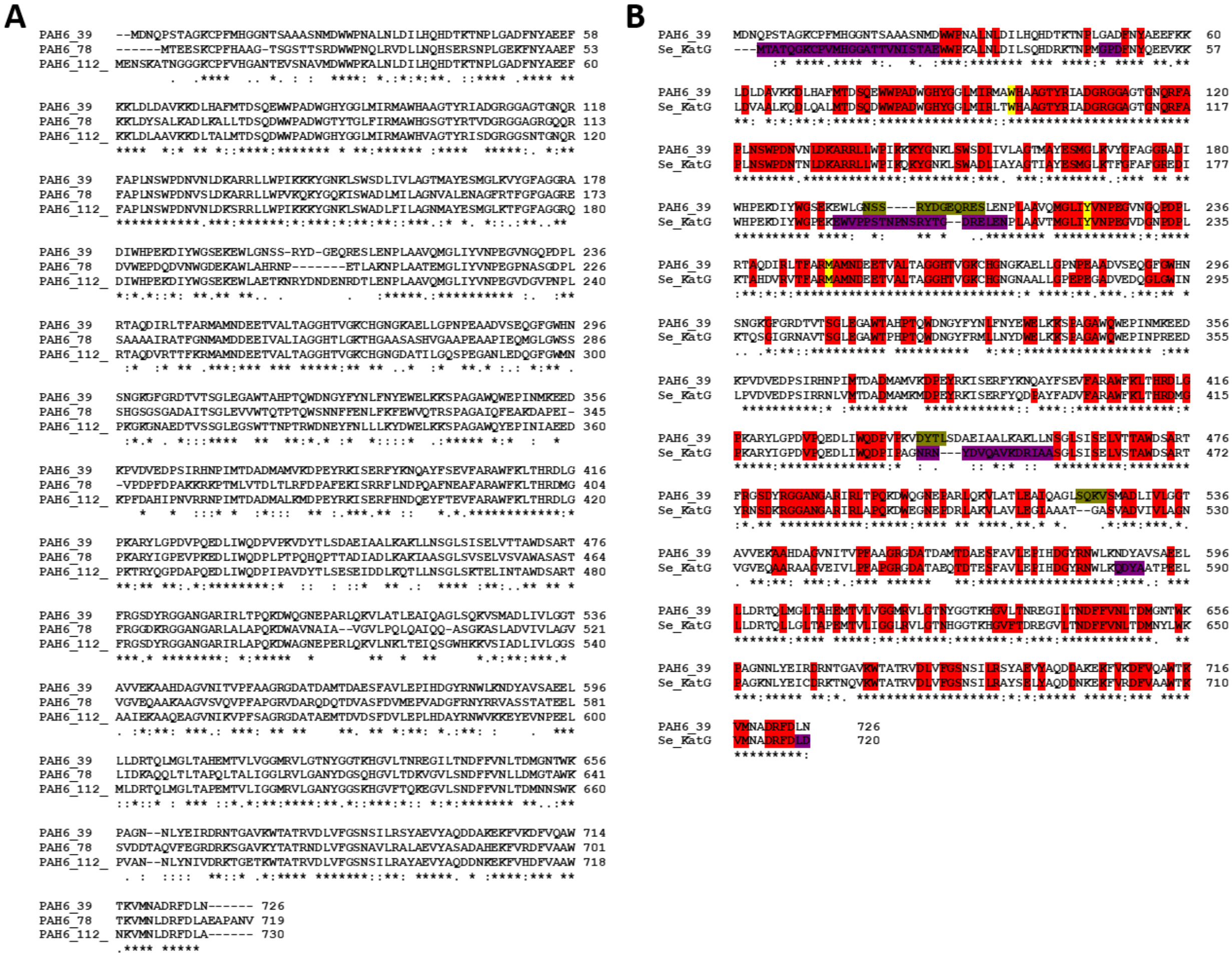

**FIGURE 4 Catalase-peroxidase sequence alignments.**

**Panel A** Multiple sequence alignment of the three catalase-peroxydase amino acid sequences which were found during metagenomic search.

**Panel B** Amino acid sequence aligment of the selected catalase-peroxydase PAH6_39 and *Synechococcus elongatus* catalase-peroxidase. Highly conserved amino acids based on literature (Kamachi *et al.*, 2015) 19abelled by red. The residues of the catalytic Met-Tyr-Trp adduct are highlighted in yellow. The flexible (loop) regions of the template were highlighted by purple, and the main differences in the three-dimensional structure between the template and the model were highlighted by ocher.

The three-dimensional fold predicted with the SWISS-MODEL server is shown in Figure 5 (homodimers and one subunit, respectively, PAH6_39: blue, template: yellow). The predicted fold of the metagenomic hit is identical to the template with minor differences observed at one loop position (a flexible loop in the template sequence – show on Figure 4 and 5B).

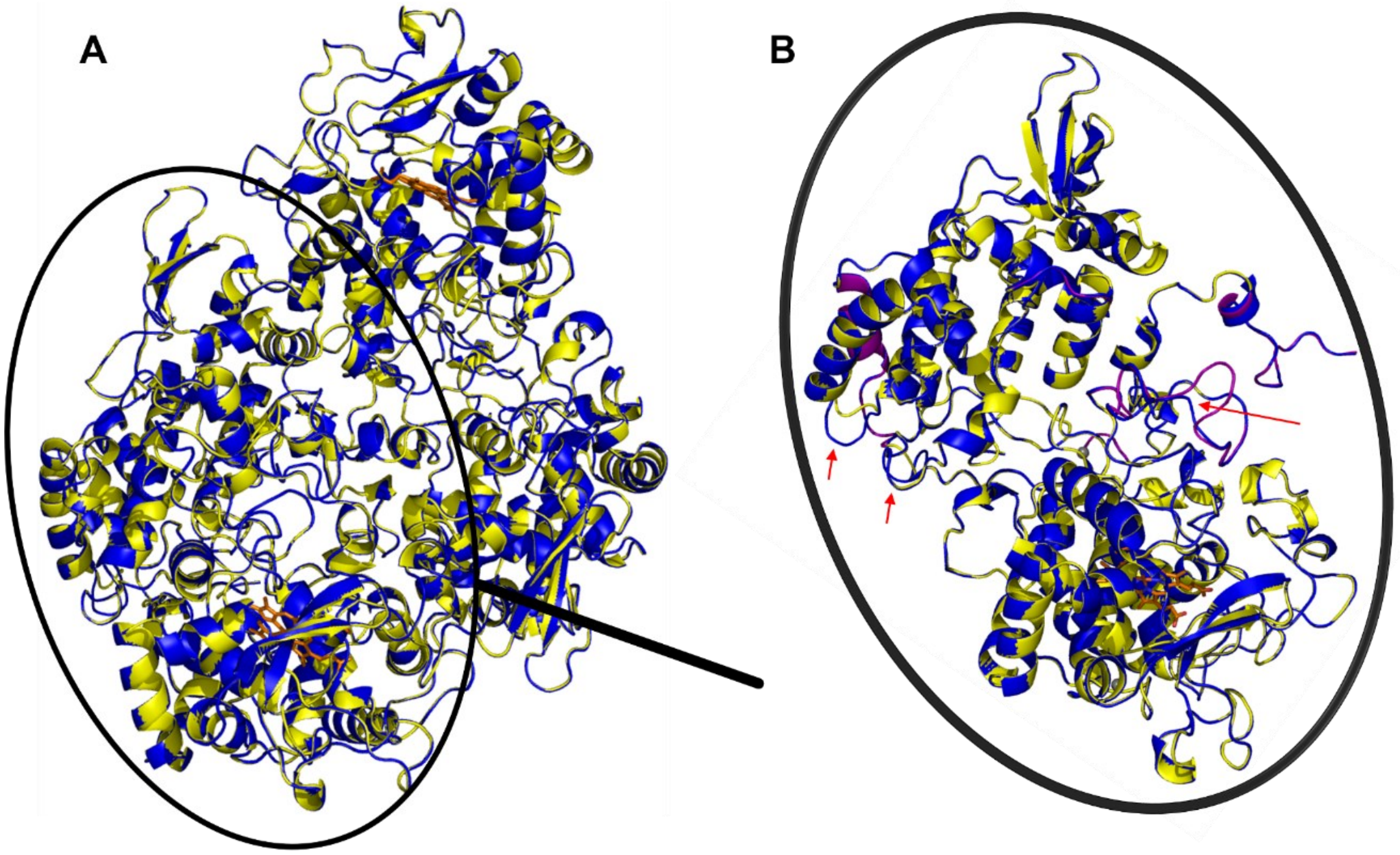

**FIGURE 5 Three-dimensional models of the catalase-peroxidase PAH6_39 built based on *Synechococcus elongatus* catalase-peroxidase (PDB ID: 3WNU)**
**Panel A** The merged homodimeric structure of the template and the model, where the PAH6_39 was labelled by blue and the *Synechococcus elongatus* catalase-peroxidase was labelled by yellow. Haems in the active sites shown in orange.
**Panel B** A close-up of the monomers. The colour scheme is same as in case of the homodimer. Na ions were labeled by gray spheres. The flexible (loop) regions of the template were colored by purple. Red arrows point on the main differences in the three-dimensional structure between the template and the models (the corresponding sequences are highlighted in ocher yellow on Fig.4B).

Based on the bioinformatic analysis and the structural modelling, we have expressed and purified two dioxygenases (PAH1_99 and PAH1_105) and one catalase-peroxidase (PAH6_39). Following the experimental protocol details in the Methods section, we have obtained the following yields from 0.5 liter medium: 59 mg, 42 mg and 49 mg, for the enzymes PAH1_99, PAH1_105 and PAH6_39 respectively. These enzyme preparations were used in the further experiments.

### 3.4 Performance of the novel enzymes in PAH degradation

In order to test the functionality of the novel enzymes in soil samples, we set a series of microcosms experiments. Using soil samples spiked with known amounts of different PAH contaminants (12.5 mg/kg, 25.9 mg/kg, 46.2 and 52.0 mg/kg for naphthalene, phenanthrene, anthracene and pyrene, respectively, cf. Methods), the effect of the enzyme addition on the level of contaminants was determined following an incubation period of 7 days.

As shown on Figure 6A, even in the absence of the added enzymes a considerable degradation of PAH compounds was observed in all of the soil microcosms experiments, suggesting the potential of an inherent microbial activity in the soils. Addition of the novel enzymes (PAH1_99, PAH1_105 and PAH6_39), predicted from metagenome searches and sequence alignments, led to further significant increase in PAH degradation only for naphthalene and phenanthrene. Anthracene and pyrene degradation was not increased by the novel enzymes in this setup.

The suggestion for the inherent microbial activity on the soils was reinforced by the Biolog EcoPlate™ test (cf. Figure 6B) where the various metabolic patterns were estimated with the AWCD values (Wolińska *et al.*, 2018). Some significant increase in the AWCD values upon addition of the novel enzyme argue for the positive effect of the enzyme proteins on the microbes in the soil samples.

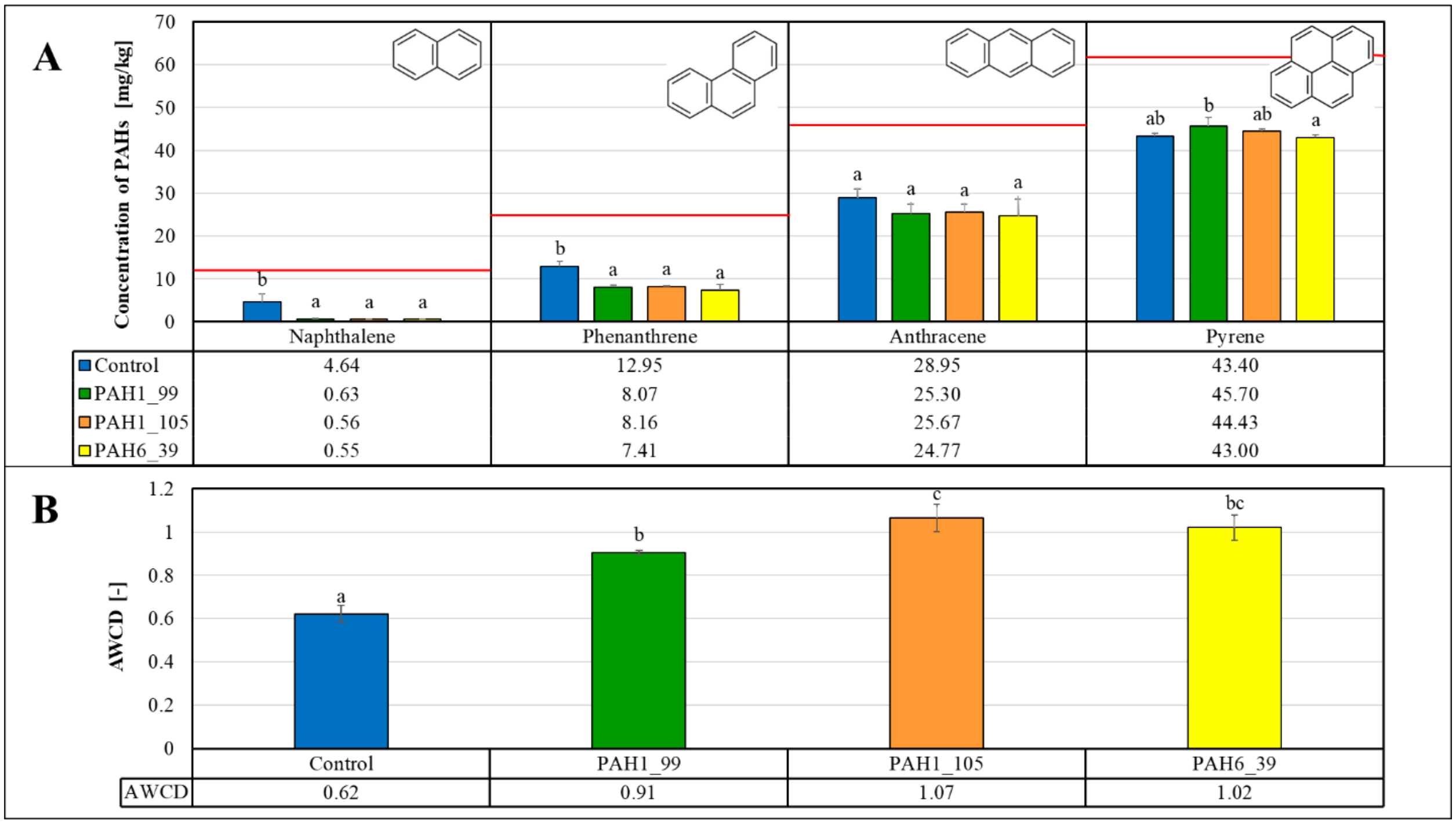

**FIGURE 6 PAH degradation (Panel A) and microbiological activity (Panel B) in the soils treated with the novel enzymes**

Data represent averages of three replicates and error bars are standard deviations. Red lines show the initial concentrations of the specific PAH compounds. Letters on the columns indicate significant differences ($p<0.05$), data in columns with different letters are statistically different, data in columns with the same letters are not statistically different.

In these experiments, we could establish significant remediation activity of the novel enzymes for some, but not all PAHs tested. It was also observed that enzyme protein addition increased the inherent microbial activity. Hence, it was not straightforward to decide whether the positive effect of the novel enzymes is due to their specific catalytic activities or to their potential to stimulate the inherent microbial activity as carbon and nitrogen source.

Towards further insights, we set up another set of experiments to investigate if combining enzyme addition with a simple inorganic oxidant molecule may increase the beneficial effects on PAH degradation. We selected $CaO_2$ for these experiments since this compound compares preferably to other oxidizing agents in terms of stability and cost efficiency.

Despite the fact that, according to other studies (Khodaveisi *et al.*, 2011; Małachowska-Jutsz & Niesler, 2015) calcium peroxide proved to be extremely effective in removing petroleum hydrocarbons and polycyclic aromatic hydrocarbons from the soil, the results of our microcosm experiments displayed otherwise in some cases. Figure 7A shows that addition of $CaO_2$ on its own actually increases the PAH levels as compared to the values observed in non-treated soils. This effect is most probably due to destroying the soil-inherent microbial flora as the metabolic activity observed in the non-treated soil is fully erased upon addition of $CaO_2$ (Figure 7B).

However, a combination of the inorganic oxidant $CaO_2$ with the novel enzymes displayed increased successful degradation of the more resistant PAHs anthracene and pyrene compounds tested in our microcosms experiments (Figure 7A). Anthracene levels drop by approx. 57–70%, while pyrene levels are decreased by approx. 56–66% upon the combined treatment of the soils with $CaO_2$ and the novel enzymes. This result also strongly suggests that the novel enzymes can provide beneficial PAH degradation in the absence of a soil-inherent microbial activity. In agreement with these findings, Wang and co-workers (Wang *et al.*, 2014) reported similar results showing that sterilization upon peroxide addition increased the degradation of pyrene because of the removal of competition from indigenous microbes.

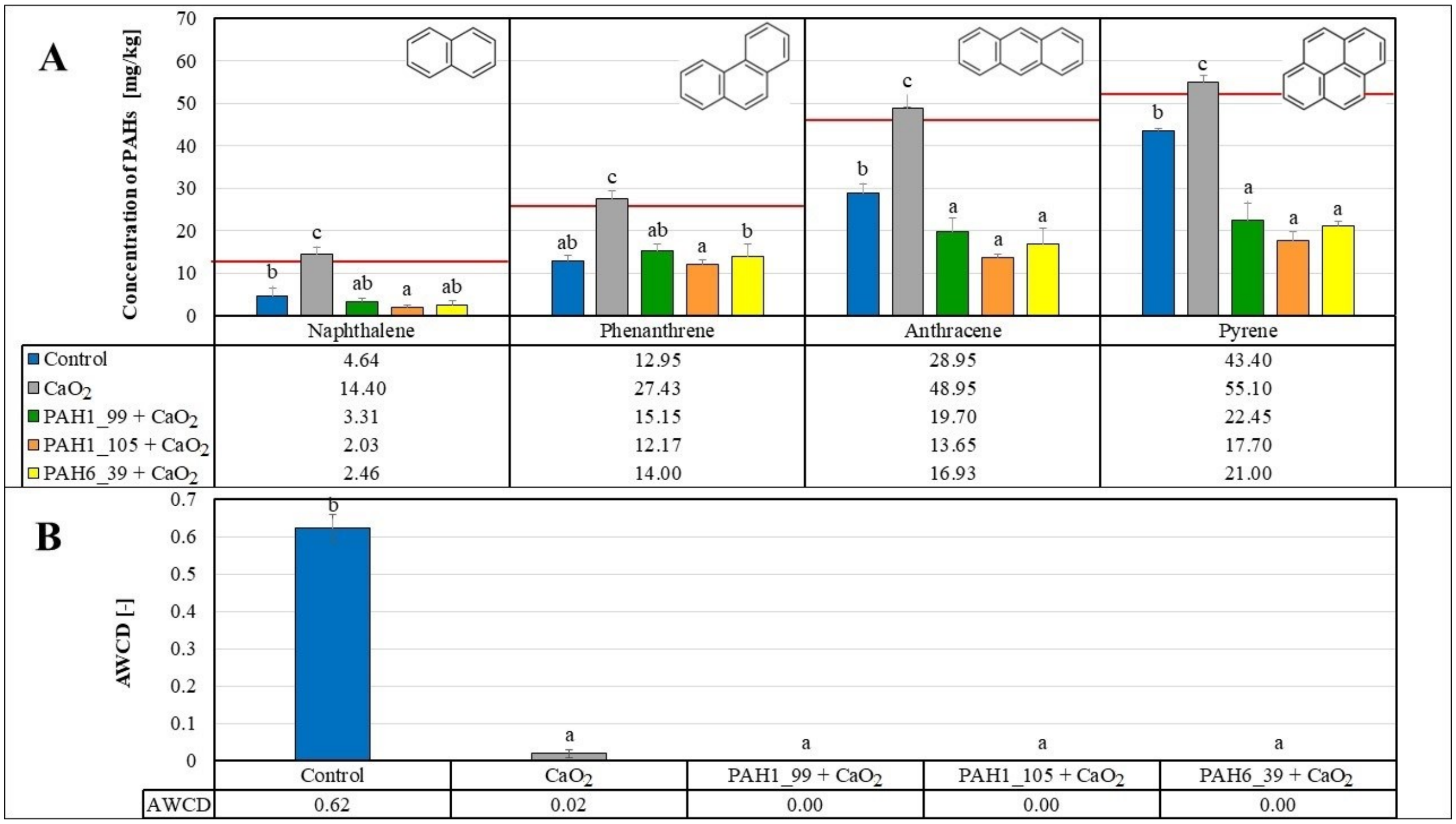

| | Naphthalene | Phenanthrene | Anthracene | Pyrene |
|---|---|---|---|---|
| Control | 4.64 | 12.95 | 28.95 | 43.40 |
| $CaO_2$ | 14.40 | 27.43 | 48.95 | 55.10 |
| PAH1_99 + $CaO_2$ | 3.31 | 15.15 | 19.70 | 22.45 |
| PAH1_105 + $CaO_2$ | 2.03 | 12.17 | 13.65 | 17.70 |
| PAH6_39 + $CaO_2$ | 2.46 | 14.00 | 16.93 | 21.00 |

| | Control | $CaO_2$ | PAH1_99 + $CaO_2$ | PAH1_105 + $CaO_2$ | PAH6_39 + $CaO_2$ |
|---|---|---|---|---|---|
| AWCD | 0.62 | 0.02 | 0.00 | 0.00 | 0.00 |

**FIGURE 7 PAH degradation (Panel A) and microbiological activity (Panel B) in the soils treated with $CaO_2$ and/or enzymes.**

Data represent averages of three replicates and error bars are standard deviations. Red lines show the initial concentrations of the specific PAH compounds. Letters on the columns indicate significant differences ($p<0.05$)

The degradation efficiency for four PAH pollutants (naphthalene, phenanthrene, anthracene, and pyrene) was calculated comparing to the analytically determined initial concentration values (Table 3). The results show that the microflora of the control group properly adapted, and the decreased PAH concentrations could be attributed to the biodegradation by indigenous microbes. Moreover, our results demonstrated that the degradation rate generally decreased with the increasing number of aromatic rings. Applied enzymes enhanced the degradation rate, whilst adding $CaO_2$ decreased the efficiency at PAHs associated with lower molecular masses. Remarkably, at naphthalene, the removal efficiency was around 95% upon enzyme treatments without peroxide, while at the combined application of enzymes and $CaO_2$, it reduced to 73–83%. The efficiency rate had a similar pattern in the case of phenanthrene, where it was 68–71% with the addition of enzymes and 41–53% with enzymes plus $CaO_2$ application.

On the other hand, in the case of anthracene, $CaO_2$ had a positive effect since the 44–46% efficiency rate with enzymes was further increased to 57–70% at the combined treatment. The highest favourable outcome of $CaO_2$ was detected at the pyrene removal, where the efficiency was 56–65% with enzymes and $CaO_2$, whilst the enzymes alone degraded only 12–17% of this pollutant.

TABLE 3 The average pollutant removal efficiency of the different treatments compared to the initial pollutant concentrations determined by GC MS

| Treatments | Removal [%] | | | |
|---|---|---|---|---|
| | Naphthalene | Phenanthrene | Anthracene | Pyrene |
| Control | 62.7 | 50.0 | 37.3 | 16.5 |
| $CaO_2$ | 0 | 0 | 0 | 0 |
| PAH1_99 | 94.9 | 68.8 | 45.2 | 12.0 |
| PAH1_105 | 95.5 | 68.5 | 44.4 | 14.5 |
| PAH6_39 | 95.6 | 71.4 | 46.3 | 17.2 |
| PAH1_99 + $CaO_2$ | 73.4 | 41.5 | 57.3 | 56.8 |
| PAH1_105 + $CaO_2$ | 83.7 | 53.0 | 70.4 | 65.9 |
| PAH6_39 + $CaO_2$ | 80.2 | 45.9 | 63.3 | 59.6 |

The degradation rate generally negatively correlated with the number of aromatic rings of the pollutants, so the removal efficiency was the highest for naphthalene and the lowest for pyrene without the use of peroxide. The combined administration of peroxide and enzymes was the most beneficial for highly stable pyrene containing four fused aromatic rings. Among these treatments, the PAH1_105 1,2-dioxygenase enzyme exhibited the highest effectiveness in the presence of $CaO_2$.

Furthermore, the applicability of the Biolog EcoPlate™ as a reliable tool for evaluating the activity of PAH-contaminated and remediated soils' microbial community and a necessary method to complement the chemical analysis of the contaminants was well demonstrated.

In conclusion, the applied novel enzymes effectively degraded the contaminants; the used $CaO_2$ slightly reduced the degradation rate in the case of naphthalene and phenanthrene while enhancing the removal of anthracene and pyrene. The novel enzyme-mediated bioremediation can be a feasible and efficient option in nutrient-poor contaminated soils with low biological activity.

## 4 CONCLUSIONS

Here, we demonstrated that an artificial intelligence-based method for identifying new PAH-degrading enzymes from still undescribed microorganisms was successful in *in vitro* studies. The Hidden Markov Model, prepared from known PAH degrading enzymes efficiently identified the sequence signature patterns in still unknown enzymes with PAH-degrading functionality. Cloning and expression of the novel enzymes was followed by functional microcosms experiments to check the soil remediation capability of the newly identified enzymes. We found the highest pyrene removal efficiency for the PAH1_105 enzyme (59%) whose effect was significantly different from PAH1_99 and PAH6_39. This result is in agreement with the prediction from the sequence and structural alignments (Figures 2 and 3) since PAH1_105 is associated with a point mutation involved in substrate selectivity.

The combined application of oxidizing agent, such as calcium peroxide, with enzymes, was significantly advantageous in the case of poorly degradable pyrene and anthracene. Interestingly, $CaO_2$ on its own was not efficient due to its strong detrimental effect on soil-inherent microbial activity. However, enzymatic degradation and removal of the PAH components was significantly increased in combination with peroxide treatment.

## AUTHOR CONTRIBUTIONS

B.G.V., V.G. and M.M. conceptualized the research. B.G.V., V.G., K.T., B.V. and M.M. conceived and designed the study and developed the methodology. K.K.N. and I.N. carried out the experimental laboratory work. B.G.V., V.G., K.K.N., I.N., M.M., K.T., and B.V analysed experimental data, conducted computational and statistical analyses. B.G.V., V.G., K.K.N., I.N., M.M., K.T., and B.V. wrote the manuscript. All authors read and approved the manuscript.

## FUNDING

KT, BV, and VG were partially supported by the Ministry of Innovation and Technology of Hungary from the National Research, Development and Innovation Fund, financed under the ELTE TKP 2021-NKTA-62 funding scheme. KT, BV, VG and BGV were partially funded by the National Research, Development and Innovation Fund under the contract 2022-1.2.2-TÉT-IPARI-UZ-2022-00003. BGV, VG and MM were partially funded by the National Research, Development and Innovation Fund under the contract NKFIH 2017-1.3.1-VKE-2017-00013.

Research was also funded by TKP2021-EGA-02 grant, implemented with the support provided by the Ministry for Innovation and Technology of Hungary from the National Research, Development and Innovation Fund and the K135231, NKP-2018-1.2.1-NKP-2018-00005 NRDI grants to BGV.

## CONFLICT OF INTEREST STATEMENT

The authors declare no conflict of interest.

## BENEFIT-SHARING STATEMENT

Benefits from this research accrue from the sharing of our data and results on public databases as described above.

## ORCID

Kinga K. Nagy

Kristóf Takács

Imre Németh https://orcid.org/0000-0003-2412-1184

Bálint Varga

Vince Grolmusz https://orcid.org/0000-0001-9456-8876

Mónika Molnár https://orcid.org/0000-0001-5296-7924

Beáta G. Vértessy https://orcid.org/0000-0002-1288-2982

## SUPPORTING INFORMATION

Additional supporting information can be found online in the Supporting Information section at the end of this article.

# Supplementary material

# 5 PAH-DEGRADING ENZYME FAMILIES USED FOR HMM PROFILE PREPARATION

## 5.1Table S1: Ring cleavage dioxygenases

| ID | Accession | Alternate Acc. | Function | Species |
|---|---|---|---|---|
| 1p1 | B0BK98 | B0BK98_9MICC | Cupin domain-containing protein | Pseudarthrobacter phenanthrenivorans |
| 1p2 | B0BK99 | B0BK99_9MICC | Dioxygenase | Pseudarthrobacter phenanthrenivorans |
| 1p3 | O24721 | PHDI_NOCSK | Dioxygenase | Nocardioides sp. |
| 1p4 | C0KUL5 | C0KUL5_9MYCO | 1-hydroxy-2-naphthoic acid dioxygenase | Mycobacterium sp. CH1 |
| 1p5 | Q51493 | NDOA_PSEAI | Naphthalene 1,2-dioxygenase system | Pseudomonas aeruginosa |
| 1p6 | Q67FT0 | Q67FT0_PSESE | Gentisate 1,2-dioxygenase | Pseudaminobacter salicylatoxidans |
| 1p7 | Q75W71 | Q75W71_9RHIZ | Gentisate 1,2-dioxygenase | Xanthobacter polyaromaticivorans |
| 1p8 | D3QJT0 | D3QJT0_ECOCB | Gentisate 1,2-dioxygenase | Escherichia coli |
| 1p9 | B1K9G3 | B1K9G3_BURCC | Gentisate 1,2-dioxygenase | Burkholderia cenocepacia |
| 1p10 | YP_004232675.1 | WP_013592695 | Gentisate 1,2-dioxygenase | Acidovorax_avenae |

## 5.2 Table S2: Bacterial peroxidases

| ID | Accession | Alternate Acc. | Function | Species |
|---|---|---|---|---|
| 2p1 | Q9R2E9 | KATG_MYCVN | Catalase-peroxidase | Mycobacterium vanbaalenii |
| 2p2 | S5TTY5 | S5TTY5_9GAMM | Catalase-peroxidase | Cycloclasticus zancles |
| 2p3 | A0A2A9KHU3 | A0A2A9KHU3_9BURK | Catalase | Collimonas sp. |

## 5.3 Table S3: Other known bacterial PAH degrading enzymes

| ID | Accession | Alternate Acc. | Function | Species |
|---|---|---|---|---|
| 3p1 | P0A110 | NDOB_PSEPU | Naphthalene 1,2-dioxygenase system | Pseudomonas putida |
| 3p2 | Q9WXG7 | DPDD_ALCFA | Cis-3,4-dihydrophenanthrene-3,4-diol dehydrogenase | Alcaligenes faecalis |
| 3p3 | Q79EM8 | PHDJ_NOCSK | Trans-2'-carboxybenzalpyruvate hydratase-aldolase | Nocardioides sp. |
| 3p4 | Q52126 | NDOR_PSEPU | Naphthalene 1,2-dioxygenase system | Pseudomonas putida |
| 3p5 | P74308 | AKR_SYNY3 | Aldo/keto reductase | Synechocystis sp. |

## 5.4 Table S4: Bacterial ring-hydroxylating oxygenases

| ID | Accession | Alternate Acc. | Function | Species |
|---|---|---|---|---|
| 4p1 | O87616 | O87616_PSEAI | Putative regulatory protein for oxygenase | Pseudomonas aeruginosa |
| 4p2 | A0A0Q8E065 | A0A0Q8E065_9ACTN | Ring-hydroxylating oxygenase | Nocardioides sp. |
| 4p3 | B3RBC6 | B3RBC6_CUPTR | Subunit of multicomponent oxygenase | Cupriavidus taiwanensis |
| 4p4 | G9G360 | G9G360_9BACT | Ring-hydroxylating dioxygenase | uncultured bacterium |
| 4p5 | G9G301 | G9G301_9BACT | dioxygenase alpha subunit | uncultured bacterium |
| 4p6 | H2IMD2 | H2IMD2_VIBSJ | ring-hydroxylating dioxygenase | Vibrio sp. |
| 4p7 | A1K8H8 | A1K8H8_AZOSB | Putative ring hydroxylating | Azoarcus sp. |

## 5.5 Table S5 Bacterial fluorene metabolism

| ID | Accession | Alternate Acc. | Function | Species |
|---|---|---|---|---|
| 5p1 | Q93UV4 | FLNB_TERSD | Fluoren-9-ol dehydrogenase | Terrabacter sp |
| 5p2 | P77567 | NHOA_ECOLI | N-hydroxyarylamine O-acetyltransferase | Escherichia coli |

## 5.6 Table S6 Bacterial phenanthrene degrading enzymes

| ID | Accession | Alternate Acc. | Function | Species |
|---|---|---|---|---|
| 6p1 | G4WYQ4 | G4WYQ4_9SPHN | Phenanthrene dioxygenase | Novosphingobium sp. |
| 6p2 | M4VW12 | M4VW12_9BURK | Phenanthrene dioxygenase | Burkholderia sp. |
| 6p3 | Q9WXG7 | DPDD_ALCFA | Cis-3,4-dihydrophenanthrene-3,4-diol dehydrogenase | Alcaligenes faecalis |
| 6p4 | Q79EM8 | PHDJ_NOCSK | Trans-2'-carboxybenzalpyruvate hydratase-aldolase | Nocardioides sp. |
| 6p5 | Q79EM7 | PHDK_NOCSK | 2-formylbenzoate dehydrogenase | Nocardioides sp. |
| 6p6 | P0A110 | NDOB_PSEPU | Naphthalene 1,2-dioxygenase system | Pseudomonas putida |
| 6p7 | Q8G8B6 | CARAA_PSERE | Carbazole 1,9a-dioxygenase | Pseudomonas resinovorans |

## 5.7 Table S7 Amino acid sequences of the hits

| Identifier | Protein sequences | Lenght (AS) | BLAST first hit | % match with BLAST's first hit |
|---|---|---|---|---|
| PAH1_16 | MEAVTKTPEREAFYKKIDGENLTALWTVMSDLITPEPKSACRPHLWKFDVIRDYMREAGKLITAKEAERRVLVLENPGLRGQSRITTSLYAGVQMVVPGDVAPAHRHSQSALRFVLEGKGAHTAVD<br>GERTAMEPGDFIITPSMTWHDHSNETDQPMFWLDGLDIPLVQFFDCSFAEGSKEDQQTITKPAGDSFARYGHNLLPVDVKRSSKTSPIFSYPYAHTREALEKARASEEWDACHGLKLKFSNPETGDF<br>AMPTIGTFIQLLPKGFKTARYRSTDATVFCPIEGHGRSRIGDAVFEWGPRDLFVVPSWQWVTHEAEDDAVLFSFSDRPVQQKLDLFREDRGNA | 346 | gentisate 1,2-dioxygenase [Bradyrhizobium sp. S23321] | 98.00% |
| PAH1_17 | MEAVQKTPEREAFYKKIDGENLSALWNVMGDLITPEPKSACRPHLWKFDAIRDYMTEAGKLITAKEAERRGLVLENPGLRGQSKITTSLFAGVQMVVPGDIAPAHRHSQSALRFVLEGKGAYTAVD<br>GERTAMEPGDFVITPSMTWHDHSNETSEPMFWLDGLDIPMVQFFDASFAEGSNEDQQKITRPAGDSFARYGHNLLPVDEKRTSKTSPIFNYPYSYTREALEQAKTRNEWDACHGLKLKFSNPETG<br>DFAMPTIGTFIQLLPKGFKTARYRSTDATVFAAIEGRGRTRIGEQTFEWGPRDLFVVPSWQWVTHEADADSVLFSFSDRPVQQKLDLFREDRGNA | 346 | gentisate 1,2-dioxygenase [Bradyrhizobium lablabi] | 99.00% |
| PAH1_99 | MKNDLIPSPVRLHAVAGHGQPDPTPELEQLYRGFEEELLVPLWTEIGDLMPRQPKSKAVPHVWRWERLKALAAQAGEIVPVGRGGERRAIALANPALGGRPFATPTLWAAIQYLMPGEDAPEHR<br>HTQHAFRFVVEGDGVWTVVNGDAVRMSRGDFLPQAGWNWHAHHNAATAPMAWIDGLDIPFSYYSESQFFEVGRDKISQAERTTAERSYSERLWAHPGLRPVSSTAATAATPLLAYRWVDTDR<br>ALADQLALEDEGQAGTLSHGHAAVRFTNPTTGGDVLPTMRCEMHRIRAGGKTKTTREVGSSVYQVFDGEGIVTVGDRTWQVTRGDLFVVPSWASFAVNALEASNLDLFRFGDAPIFDALHNYRT<br>EIIS | 374 | cupin domain-containing protein [Acidovorax sp. OV235] | 86.00% |
| PAH1_102 | MSNHDGFQQAPVHNAMAPDDSPELRQLYADFEAGHMMPLWTQIGNLMPKHPMPRAVPHVWKWSDLYPLAKRSGDLVPVGRGGERRAIGLGNPGLEGRPYISPTLWCAIQYLGPRETAPEHR<br>HAQNAFRFVIEGEGVWTVVNGDPVRMSRGDLLLTPGWNFHGHQNVTDKPMAWIDGLDIPFSYQNDVGFFEFGSENLTDITTPQYSRGERLWCHPGLRPLSGLANTVSSPIGAYRWEHTDRALD<br>EQLRLEEEGFPGVQEKGHAAVRFINPTTGGDIMSSIRAEFHRLRAGAVTAERREVGSRVFQVFEGRGQVMLDGVTRHLEKGDLFVVPSWISWSLQAESQFDLFSFSDAPIMADVDENRLIVAHTLK<br>SWQVAGQSRASAYGLAAPERPIRPTGH | 397 | cupin domain-containing protein [Pseudomonas sp. GM55] | 96.00% |
| PAH1_105 | MNTATLRAAPPQADERRAYYERIRPLNLTPLWESLHALVPREPQTPCVPALWRYDDIRPLLMESAELITADEAVRRVLVLENPAIPGRSSITQSLYAGLQLIMPGEVAPSHRHVQSALRFIVDGKGAYT<br>TVGGERTTMHPGDFIITPSWAWHDHGNEGIEGVSEPVVWLDGLDIPMVRFFDAGFAENAEAKVQHVARPEGHSLARFGHNMVPVRHDHTSATSPIFNYPYLRSREALAQLQMQEAPDAWLGH<br>KLRYINPLTGGSPMPTIATNLQLLPRGFAGKTHRMTDGAVYSVVEGRGHADIGGQRFDFGPRDTFVVPSWAPLKLVASDDVVLFSFSDRPVQQAMGVLREAFLED | 356 | gentisate 1,2-dioxygenase [Variovorax sp. YR216] | 91.00% |
| PAH1_117 | MPVIPNPKAVPHVWKWSRLYPLAERSGDLVPVGRGGERRAIGLSNPGLGGRAYVSPTLWAAIQYLGPRETAPEHRHAQNAFRFVVEGEGVWTVVNGDPVRMSRGDLLLTPGWNFHGHHNDT<br>DHPMAWIDGLDIPFSYQNDVGFFEFGSDRVTDYATPQFSRGERLWAHPGLRPLSQLTDTVSSPLAAYRWEFTDRALTEQLLLEDEGQPATVGQGHAAIRYVNPTTGGDVMPTIRCEFHRLREGTV<br>TPPRREVGSSVFQVFEGTGSVVLNGTETKLEKGDMFVVPSWVACSLQAETRFDLFRFSDAPIIERLGFARTLVENNER | 324 | cupin domain-containing protein [Mesorhizobium australicum] | 89.00% |
| PAH6_39 | MDNQPSTAGKCPFMHGGNTSAAASNMDWWPNALNLDILHQHDTKTNPLGADFNYAEEFKKLDLDAVKKDLHAFMTDSQEWWPADWGHYGGLMIRMAWHAAGTYRIADGRGGAGTGNQ<br>RFAPLNSWPDNVNLDKARRLLWPIKKKYGNKLSWSDLIVLAGTMAYESMGLKVYGFAGGRADIWHPEKDIYWGSEKEWLGNSSRYDGEQRESLENPLAAVQMGLIYVNPEGVNGQPDPLRTAQ<br>DIRLTFARMAMNDEETVALTAGGHTVGKCHGNGKAELLGPNPEAADVSEQGFGWHNSNGKGFGRDTVTSGLEGAWTAHPTQWDNGYFYNLFNYEWELKKSPAGAWQWEPINMKEEDKPV<br>DVEDPSIRHNPIMTDADMAMVKDPEYRKISERFYKNQAYFSEVFARAWFKLTHRDLGPKARYLGPDVPQEDLIWQDPVPKVDYTLSDAEIAALKAKLLNSGLSISELVTTAWDSARTFRGSDYRGGA<br>NGARIRLTPQKDWQGNEPARLQKVLATLEAIQAGLSQKVSMADLIVLGGTAVVEKAAHDAGVNITVPFAAGRGDATDAMTDAESFAVLEPIHDGYRNWLKNDYAVSAEELLLDRTQLMGLTAHE<br>MTVLVGGMRVLGTNYGGTKHGVLTNREGILTNDFFVNLTDMGNTWKPAGNNLYEIRDRNTGAVKWTATRVDLVFGSNSILRSYAEVYAQDDAKEKFVKDFVQAWTKVMNADRFDLN | 726 | catalase/peroxidase HPI [Methylotenera sp. 24-45-7] | 87.00% |
| PAH6_78 | MTEESKCPFHAAGTSGSTTSRDWWPNQLRVDLLNQHSERSNPLGEKFNYAAEFKKLDYSALKADLKALLTDSQDWWPADWGTYTGLFIRMAWHGSGTYRTVDGRGGAGRGQQRFAPLNSWP<br>DNVSLDKARRLLWPVKQKYGQKISWADLMILAGNVALENAGFRTFGFGAGREDVWEPDQDVNWGDEKAWLAHRNPETLAKNPLAATEMGLIYVNPEGPNASGDPLSAAAAIRATFGNMAMD<br>DEEIVALIAGGHTLGKTHGAASASHVGAAPEAAPIEQMGLGWSSSHGSGSGADAITSGLEVVWTQTPTQWSNNFFENLFKFEWVQTRSPAGAIQFEAKDAPEIVPDPFDPAKKRKPTMLVTDLTL<br>RFDPAFEKISRRFLNDPQAFNEAFARAWFKLTHRDMGPKARYIGPEVPKEDLIWQDPLPTPQHQPTTADIADLKAKIAASGLSVSELVSVAWASASTFRGGDKRGGANGARLALAPQKDWAVNAI<br>AVGVLPQLQAIQQASGKASLADVIVLAGVVGVEQAAKAAGVSVQVPFAPGRVDARQDQTDVASFDVMEPVADGFRNYRRVASSTATEELLIDKAQQLTLTAPQLTALIGGLRVLGANYDGSQHG<br>VLTDKVGVLSNDFFVNLLDMGTAWKSVDDTAQVFEGRDRKSGAVKYTATRNDLVFGSNAVLRALAEVYASADAHEKFVRDFVAAWTKVMNLDRFDLAEAPANV | 719 | MULTISPECIES: catalase/peroxidase HPI [Acidovorax] | 99.00% |
| PAH6_112 | MENSKATNGGGKCPFVHGANTEVSNAVMDWWPKALNLDILHQHDTKTNPLGADFNYAEEFKKLDLAAVKKDLTALMTDSQDWWPADWGHYGGLMIRMAWHVAGTYRISDGRGGSNTGN<br>QRFAPLNSWPDNVNLDKSRRLLWPIKKKYGNKLSWADLFILAGNMAYESMGLKTFGFAGGRQDIWHPEKDIYWGSEKEWLAETKNRYDNDENRDTLENPLAAVQMGLIYVNPEGVDGVPNPLR<br>TAQDVRTTFKRMAMNDEETVALTAGGHTVGKCHGNGDATILGQSPEGANLEDQGFGWMNPKGKGNAEDTVSSGLEGSWTTNPTRWDNEYFNLLLKYDWELKKSPAGAWQYEPINIAEEDKP<br>FDAHIPNVRRNPIMTDADMALKMDPEYRKISERFHNDQEYFTEVFARAWFKLTHRDLGPKTRYQGPDAPQEDLIWQDPIPAVDYTLSESEIDDLKQTLLNSGLSKTELINTAWDSARTFRGSDYRG<br>GANGARIRLAPQKDWAGNEPERLQKVLNKLTEIQSGWHKKVSIADLIVLGGSAAIEKAAQEAGVNIKVPFSAGRGDATAEMTDVDSFDVLEPLHDAYRNWVKKEYEVNPEELMLDRTQLMGLTA<br>PEMTVLIGGMRVLGANYGGSKHGVFTQKEGVLSNDFFVNLTDMNNSWKPVANNLYNIVDRKTGETKWTATRVDLVFGSNSILRAYAEVYAQDDNKEKFVHDFVAAWNKVMNLDRFDLA | 730 | peroxidase, partial [Lutibacter sp. BRH_c52] | 94.00% |

**5.8 Figure S1 Disorder predictions of the selected proteins. A predictions for dioxygenases B predictions for catalase-peroxidases. The plot shows the IUPRED2 sequence−prediction profile indicated in red, as well as the ANCHOR2 prediction shown in blue. Residues with scores above 0.5 are predicted as disordered.**

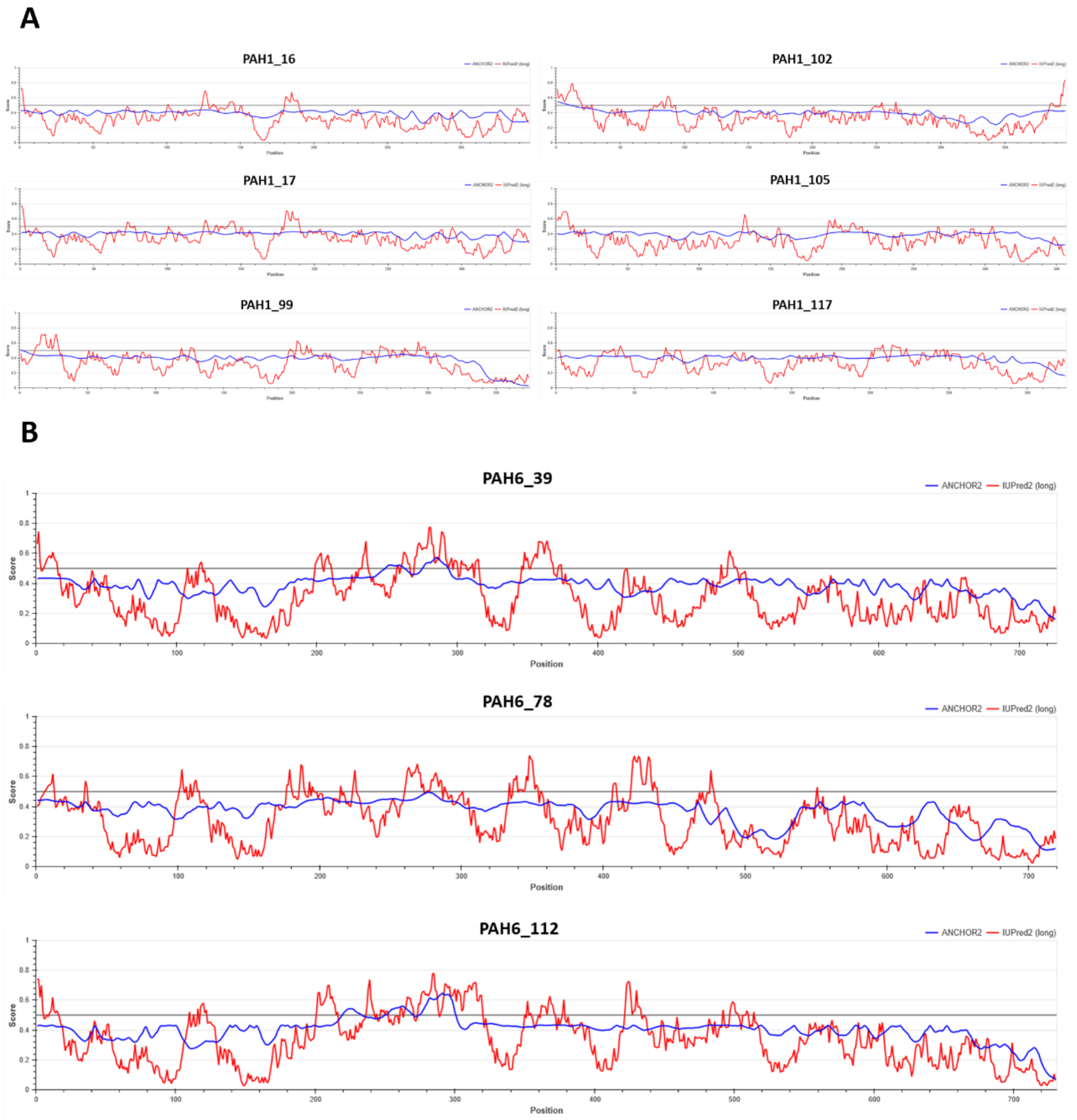